\documentclass[useAMS,usenatbib,usegraphicx,letterpaper]{mn2e}

\usepackage{multirow} 
\usepackage{epsfig}
\usepackage{subfigure}

\newcommand{\etal}{et al.}
\newcommand{\leqsim}{\raisebox{-0.6ex}{$\,\stackrel
        {\raisebox{-.2ex}{$\textstyle <$}}{\sim}\,$}}
\newcommand{\geqsim}{\raisebox{-0.6ex}{$\,\stackrel
        {\raisebox{-.2ex}{$\textstyle >$}}{\sim}\,$}}

\def\mnras{MNRAS}
\def\nat{Nature}
\def\apj{ApJ}
\def\araa{Annual Review of Astronomy \& Astrophysics}
\def\apjs{ApJS}
\def\apjl{ApJ}
\def\aap{A\&A}
\def\aj{AJ}

\def\procspie{Proc. SPIE}
\def\pasp{Proc. of the Astronomical Society of the Pacific}
\def\pasa{Publications of the Astronomical Society of Australia}

\title[A high $\lambda$, true Seyfert~2 galaxy candidate] {A high
  Eddington--ratio, true Seyfert~2 galaxy candidate: implications for
  broad--line--region models}

\author[G.\ Miniutti \etal]
       {\parbox{\textwidth}{G.~Miniutti,$^{1}$\thanks{E-mail:
             \texttt{gminiutti@cab.inta-csic.es}} R.~D.~Saxton,$^{2}$
           P.~M.~Rodr\'{i}guez--Pascual,$^{2}$, A.~M.~Read,$^{3}$
           P.~Esquej,$^{1,4,5}$ M.~Colless,$^{6}$ P.~Dobbie$^{6,7}$ and
           M.~Spolaor$^{6}$
}\vspace{0.5cm}\\
\parbox{\textwidth}{
$^{1}$Centro de Astrobiolog\'{i}a (CSIC--INTA), Dep. de Astrof\'{i}sica; 
ESAC, PO Box 78, E-28691, Villanueva de la Ca\~nada, Madrid, Spain
\\
$^{2}$XMM SOC, ESAC, P.O. Box 78, E-28691, Villanueva de la Ca\~nada, Madrid, Spain\\
$^{3}$Dept. of Physics and Astronomy, University of Leicester, Leicester LE1~7RH, UK
\\
$^{4}$Instituto de F\'{\i}sica de Cantabria, CSIC-Universidad de Cantabria, 39005 Santander, Spain
\\
$^{5}$Departamento de F\'{\i}sica Moderna, Universidad de Cantabria, Avda. de Los Castros s/n, 39005 Santander, Spain
\\
$^{6}$Australian Astronomical Observatory, PO Box 915, North Ryde, NSW 1670, Australia\\
$^{7}$School of Maths \& Physics, University of Tasmania, Sandy Bay, 7001, Australia
}}

\pagerange{\pageref{firstpage}--\pageref{lastpage}}
\pubyear{2012}

\usepackage{times}

\begin{document}

\label{firstpage}

\maketitle

\begin{abstract}

A bright, soft X--ray source was detected on 2010 July 14 during an
{\it XMM--Newton} slew at a position consistent with the galaxy
GSN~069 ($z=0.018$). Previous {\it ROSAT} observations failed to
detect the source and imply that GSN~069 is now $\geq$~240 times
brighter than it was in 1994 in the soft X--ray band. Optical spectra
(from 2001 and 2003) are dominated by unresolved emission lines with
no broad components, classifying GSN~069 as a Seyfert~2 galaxy. We
report here results from a $\sim$~1~yr monitoring with {\it Swift} and
{\it XMM--Newton}, as well as from new optical spectroscopy. GSN~069
is an unabsorbed, ultra--soft source in X--rays, with no flux detected
above $\sim$1~keV. The soft X--rays exhibit significant variability
down to timescales of hundreds of seconds. The UV--to--X--ray spectrum
of GSN~069 is consistent with a pure accretion disc model which
implies an Eddington ratio $\lambda \simeq 0.5$ and a black hole mass
of $\simeq 1.2\times 10^6~M_\odot$. A new optical spectrum, obtained
$\sim$~3.5~months after the {\it XMM--Newton} slew detection, is
consistent with earlier spectra and lacks any broad line
component. The lack of cold X--ray absorption and the short timescale
variability in the soft X--rays rule out a standard Seyfert~2
interpretation of the source. The present Eddington ratio of GSN~069
exceeds the critical value below which no emitting broad--line--region
(BLR) forms, according to popular models, so that GSN~069 can be
classified as a bona--fide high Eddington--ratio true Seyfert~2
galaxy. We discuss our results within the framework of two possible
scenarios for the BLR in AGN, namely the two--phase model (cold BLR
clouds in pressure equilibrium with a hotter medium), and models in
which the BLR is part of an outflow, or disc--wind. Finally, we point
out that GSN~069 may be a member of a population of super--soft AGN
whose SED is completely dominated by accretion disc emission, as it is
the case in some black hole X--ray binary transients during their
outburst evolution. The disc emission for a typical AGN with black
hole mass of $10^7-10^8~M_\odot$ does not enters the soft X--ray band,
so that GSN~069--like objects with larger black hole mass (i.e. the
bulk of the AGN population) are missed by current X--ray surveys, or
mis--classified as Compton--thick candidates. If the analogy between
black hole X--ray binary transients and AGN holds, the lifetime of
these super--soft states in AGN may be longer than $10^4$ years,
implying that the actual population of super--soft AGN may not be
negligible, possibly contaminating the estimated fraction of
heavily obscured AGN from current X--ray surveys.

\end{abstract}


\section{Introduction}

One of the key ideas upon which we base our understanding of Active
Galactic Nuclei (AGN) is that type 1 and type 2 AGN have no intrinsic
physical differences, their classification being dominated by the
presence/absence of absorbing material in our line--of--sight, which
is considered to be orientation--dependent (Antonucci 1993; see also
Elitzur 2012). Although this Unified Model has been extremely
successful, additional ingredients are likely needed to account
for some observational facts that apparently are in conflict with the
expectations. Among these, it is well known that a significant
fraction of the brightest Seyfert~2 galaxies lack broad optical lines
even in polarised light (e.g. Tran 2003). Moreover, a (still
  relatively small) sample of Seyfert~2 galaxies have been found to be
  unabsorbed in the X--rays, thus challenging the Unified Model (Pappa
  et al. 2001; Panessa \& Bassani 2002; Brightman \& Nandra 2008;
  Panessa et al. 2009; Shi et al. 2010; Bianchi et al. 2012).

Various explanations have been proposed to solve the apparent puzzle
of unabsorbed Seyfert~2 galaxies. The proposed explanations range from
dilution (the broad, optical  lines can be overwhelmed by the host galaxy
contribution), to large amplitude, long--term variability (either in
luminosity or in absorption properties), and signal--to--noise
issues. However, good signal--to--noise optical and
quasi--simultaneous X--ray data seem to point towards the real
existence of a population of genuine unabsorbed Seyfert~2 galaxies
(e.g. Panessa et al. 2009; Bianchi et al. 2012). At present, some of
the best unabsorbed Seyfert~2 candidates which have been
quasi--simultaneously observed in the optical and X--rays are
NGC~3147, NGC~3660 and Q~2131--427 (Bianchi et al. 2012).

From an observational point of view, unabsorbed Seyfert~2
(and Seyfert~2 without hidden BLR, HBLR) galaxies are predominantly found at
low Bolometric luminosity or, correspondingly, low Eddington ratio
(Bian \& Gu 2007; Panessa et al. 2009; Shi et al. 2010; Wu et
al. 2011; Marinucci et al. 2012). This is generally interpreted via
theoretical models in which the broad line region (BLR) is part of an
outflow (e.g. Emmering, Blandford \& Shlosman 1992; Murray et
al. 1995; Nicastro 2000; Elitzur \& Ho 2009; Trump et al. 2011). For
instance, Nicastro (2000) proposes that the BLR originates in a disc--wind
which is quenched below the critical mass accretion rate at
which the accretion disc is gas--pressure--dominated throughout
($\dot{m}_{\rm{crit}} \sim 2.4\times 10^{-3}$ in Eddington units and
for typical radiative efficiency $\eta =0.06$, viscosity parameter
$\alpha = 0.1$, and black hole mass of $10^8~M_\odot$). Although
unobscured/non--HBLR Seyfert~2 are indeed observationally confirmed to
have accretion rates lower than the critical one (Trump et al. 2011;
Marinucci et al. 2012; Bianchi et al. 2012), the mere existence of
many type 1 Seyfert galaxies with clear BLR optical lines at much
lower accretion rates represents a challenge for such model
(e.g. M~81, Peimbert \& Torres--Peimbert 1981; Filippenko \& Sargent
1988; Ho 2008; Elitzur \& Ho 2009). Indeed, the outflow model
for the BLR/torus formation proposed by Elitzur \& Ho (2009) on the
basis of the original Elitzur \& Shlosman (2006) description, predicts that
the BLR should cease to exist only below $\dot{m}_{\rm{crit}} \sim
1-2\times 10^{-6}$ (for a $10^8~M_\odot$ black hole), more than 3
orders of magnitude smaller than that proposed by Nicastro
(2000). 

Wang \& Zhang (2007) have tested the Unified Model on a large sample
of 243 local Seyfert galaxies and have proposed a scenario in which
evolution plays an important role by controlling relevant physical
parameters such as black hole mass and accretion rate, obscuring torus
opening angle, and gas--to--dust ratio of the torus itself. Their
evolutionary sequence for Seyfert galaxies starts with
optically--selected Narrow--Line--Seyfert~1 galaxies associated with
high Eddington ratios and relatively small black hole masses, and ends
with unabsorbed Seyfert~2 galaxies with no HBLR, a stage that is
reached when the mass accretion rate is so low that the BLR
disappears.

On the other hand, Wang et al. (2012) propose a different evolutionary
sequence based upon the effects of star formation in AGN
self--gravitating discs. These authors propose that warm skins are
formed above a star--forming disc due to the diffusion of gas driven
by supernova explosions, and show that the system evolution implies
the episodic appearance of BLR with different properties, according to
the specific evolutionary stage. In particular, Wang et al. (2012)
predict the existence of an initial phase (phase I in their model)
which is characterised by relatively high Eddington ratio and no BLR,
as the line--emitting--region is still forming. These objects are
predicted to be rare (and, for this reason, they are called
{\it{Panda}} AGN) with a $\sim 1$ per cent probability of occurrence in
the overall population. Nevertheless, Panda AGN can in principle be
distinguished from more typical low--luminosity unabsorbed Seyfert~2
galaxies because of their much higher Eddington ratio.

Here we present results obtained from optical--to--X--ray observations
of GSN~069, a soft X--ray variable, unabsorbed, high Eddington--ratio
Seyfert~2 galaxy. We present quasi--simultaneous optical and X--ray
spectra of this peculiar source, and we discuss our observational
results in the context of models for the BLR in AGN, as well as their
application to this particular case of a radiatively--efficient
unobscured Seyfert~2 galaxy candidate.

\section{The case of GSN~069}

A relatively bright ultra--soft X--ray source was
detected on 2010--07--14 during an {\it XMM--Newton} slew from a
position consistent with the galaxy GSN~069
(a.k.a. 6dfg~0119087--341131) at redshift $z=0.0181$. {\it
  XMM--Newton} recorded an EPIC--pn count--rate of $\sim 1.5$ counts/s
in the soft 0.2--2~keV band, corresponding to a flux of $\sim
2.4\times 10^{-12}$~erg~s$^{-1}$~cm$^{-2}$, according to the best
available spectral decomposition (see Sections below). Previous ROSAT
pointed observations performed $\sim 16$~yr earlier failed to detect
the source, from which we infer that GSN~069 was at least 240 times
fainter in the soft X--rays than during the {\it XMM--Newton} slew
detection. The source was subsequently monitored with {\it SWIFT} for
about one year. In order to better characterise the X--ray spectrum
and the short--timescale variability of the source, we obtained a
further $\sim 15$~ks {\it XMM--Newton} observation on 2010--12--02
which again detected the source in the soft X--rays at a similar flux
level (Saxton et al. 2011).

\begin{figure}
\begin{center}
\includegraphics[width=0.51\textwidth,height=0.46\textwidth,angle=-0]{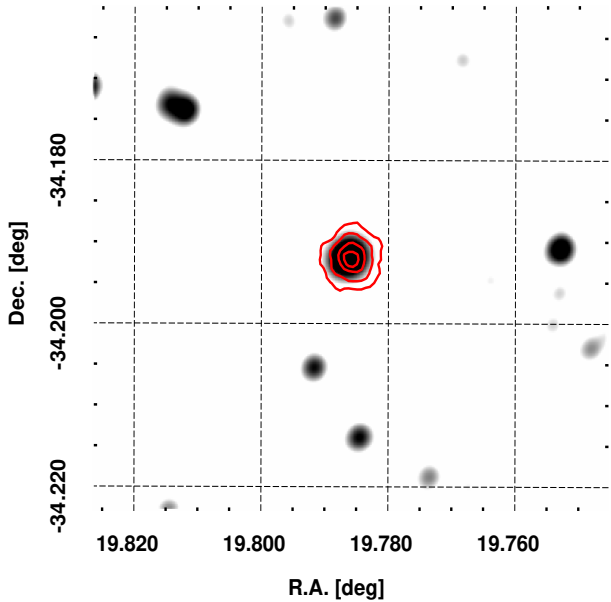}
\caption{A 4$'\times$4$'$ {\it WISE} 3--colour image centred on
  GSN~069. The contours are from the {\it XMM--Newton} pointed
  observation. The UVOT--enhanced {\it Swift} X--ray position is R.A.:
  01h19m08.66s and Dec: --34d11m30.4s with a 90 per cent error of
  1.9$''$ in radius. The {\it Swift} error--circle lies within the
  innermost {\it XMM--Newton} contour.}
\label{wiseXMM}
\end{center}
\end{figure}

GSN~069 was also detected in the far and near UV with {\it GALEX} and
is present in the {\it WISE} all-sky data release as well as in the
2MASS point--source catalogue. Fig.~\ref{wiseXMM} shows a
4$'\times$4$'$ region of the {\it WISE} 3--colour image centred on
GSN~069 together with the {\it XMM--Newton} EPIC--pn contours from the
2010 pointed observation. Two, consistent, optical spectra, taken in
2001 and 2003 in the 2dF and 6dF surveys of the Anglo--Australian
Telescope (AAT) show unresolved Balmer lines with no apparent broad
components which, together with the diagnostic line ratios, classify
the source as a Seyfert~2 galaxy. Here we report results from our
X--ray ({\it XMM--Newton} and {\it Swift}) observations of this
peculiar Seyfert~2 galaxy as well as from a new AAT optical spectrum
obtained $\sim$~3.5~months after the {\it XMM--Newton} slew
detection. We complement our work with archival data which enable us
to build the IR--to--X--ray spectral energy distribution (SED) of the
source.

Throughout the paper, we adopt a cosmology with
H$_0=70$~km~s$^{-1}$~Mpc$^{-1}$, $\Omega_\Lambda= 0.73$, and $\Omega_M
= 0.27$.

\section{Observations}

The {\it XMM--Newton} slew (9194000004), which detected GSN~069 for
the first time on 2010--07--14 with a soft X--rays count rate of
$1.5\pm 0.4$~counts~s$^{-1}$, was performed with the EPIC--pn camera
operated in Full Frame mode with the Medium optical filter
applied. GSN~069 was then the target of a pointed {\it XMM--Newton}
observation on 2010-12-02 performed in Full Frame mode with the Thin
optical filter applied. The data were reduced as standard using the
dedicated SAS v11.0 software. The final net exposures are $\sim 11$~ks
in the pn and $\sim 14$~ks in the two MOS cameras. The EPIC pn, MOS~1
and MOS~2 spectra were grouped using the {\tt specgroup} SAS task so
that i) each group has a signal--to--noise ratio $S/N \geq 4$, and ii)
no group over--samples the FWHM instrumental resolution at the central
energy of the group by more than a factor of $3$. All groups comprise
more than 20 background--subtracted counts, enabling us to use the
$\chi^2$ statistic for spectral fitting. The estimated background
counts are comparable to the source--only ones above $\sim$~0.95~keV
in the pn spectrum, and above $\sim$~0.7~keV in the MOS~1 and 2. We
then consider here the X--ray data above 0.3~keV and below the above
detector--dependent high--energy threshold. The RGS data were also
reduced, but the short exposure and relatively low flux in the soft
X--rays make them effectively meaningless for spectral
analysis. During the observation, simultaneous optical and UV data
were collected with the Optical Monitor (OM) with filters B ($\sim
4500$\AA), W1 ($\sim 2900$\AA), and M2 ($\sim 2300$\AA).

GSN~069 was monitored with 13 {\it Swift} observations performed in
standard pc mode from 2010--08--26 to 2011--08--18. The data were
reduced as standard. As none of the individual {\it Swift}--XRT
spectra is of high enough quality to perform detailed spectral
analysis, the XRT data are used here to construct the 0.5--2~keV flux
light curve of GSN~069 during the $\sim$~1~yr monitoring of the
source. The X--ray data were complemented with photometry from the
Ultraviolet/Optical Telescope (UVOT) which was operated in the
U and/or W1 filters. U and W1 magnitudes of the source were $16.7$ in
both filters with no significant variation, in agreement with
the {\it XMM--Newton} OM photometry.

We observed GSN~069 with the AAOmega 2dF spectrograph (
Saunders et al. 2004; Sharp et al. 2006) at the focus of the
Anglo--Australian Telescope on 2010--10--27 for 3$\times$600~s. We
used the 580V (3700--5800~\AA) and 385R (5600--8800~\AA)
low--resolution gratings ($R\sim 1300$). The 2dF data were reduced
using the 2dFDR software of the AAT (e.g. Sharp \& Birchall 2010)
which performs bias subtraction, fibre--flat fielding and wavelength
calibration in an automated manner. The response curves for the blue
and red arms were determined with an observation of the bright DA
white dwarf EG21 and were then used to obtain a relative flux
calibration of the spectrum of GSN~069. The spectrum was then
re--normalised to the B--band ($\sim 4500$\AA) {\it XMM--Newton} flux.

\begin{table}
\caption{GSN~069 observation dates and 0.2--2~keV fluxes. The fluxes
  have been estimated assuming a simple black body model and Galactic
  absorption fit to the pointed {\it XMM--Newton} observation. As the
  spectral model is simpler than the best--fitting one (also
  comprising significant warm absorption, see
  Section~\ref{secXrayspec}), the observed flux during the {\it
    XMM--Newton} pointed observation is about 10 per cent higher than
  that reported here. We however prefer to use the simpler spectral
  model here to ease comparison with future observations. Quoted
  uncertainties are given at the $2\sigma$ confidence level and
  are statistical only, while systematic, model--dependent
  uncertainties may be larger.}
\label{tab1}      
\begin{center}
\begin{tabular}{l c l }
\hline\hline                 
Mission & Date & Flux$^{a}$ \\
\\
RASS  & 1990                          & $<0.05$  \\
ROSAT-PSPC    &  1993-07-13  & $<0.009$   \\
ROSAT-PSPC    &  1994-06-29  & $<0.007$   \\
XMM slew  &   2010-07-14     & $2.4\pm{0.7}$   \\
SWIFT  & 2010-08-26          & $2.1\pm 0.6$   \\
SWIFT  & 2010-08-27          & $2.1\pm 0.3$ \\
SWIFT  & 2010-10-27          & $1.8\pm 0.3$ \\
SWIFT  & 2010-11-24          & $1.6\pm 0.3$ \\
XMM pointed &   2010-12-02   & $2.03\pm{0.03}$\\
SWIFT  & 2010-12-22          & $1.4\pm 0.3$\\
SWIFT  & 2011-01-19          & $1.4\pm 0.3$\\
SWIFT  & 2011-02-16          & $1.5\pm 0.3$ \\
SWIFT  & 2011-04-25          & $1.4\pm 0.3$ \\
SWIFT  & 2011-05-23          & $1.7\pm 0.3$ \\
SWIFT  & 2011-06-20          & $1.8\pm 0.3$ \\
SWIFT  & 2011-07-17          & $1.4\pm 0.3$ \\
SWIFT  & 2011-08-15         & $2.0\pm 0.3$ \\
SWIFT  & 2011-08-18          & $1.9\pm 0.5$ \\
\hline                        
\end{tabular}
\\
\end{center}
$^{a}$ Flux in the 0.2--2~keV band in units of
$10^{-12}$~erg~s$^{-1}$~cm$^{-2}$.
\end{table}

\section{The X--ray variability of GSN~069}

\begin{figure}
\begin{center}
\includegraphics[width=0.33\textwidth,height=0.45\textwidth,angle=-90]{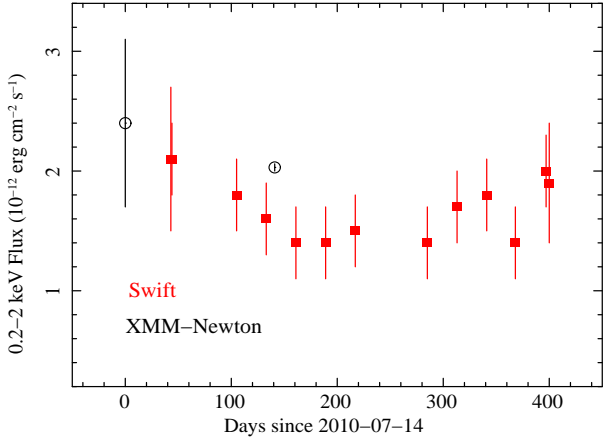}
\caption{Historical 0.2--2~keV observed flux of GSN~069 since the {\it
    XMM--Newton} slew detection on 2010-07-14. Open circles represent
  the {\it XMM-Newton} pn slew and pointed observations; filled boxes
  are from the {\it Swift}--XRT detector (see Table~\ref{tab1}).}
\label{histlc}
\end{center}
\end{figure}

All available X--ray observation dates and soft X--ray fluxes are
reported in Table~\ref{tab1}. The historical X-ray light curve of
GSN~069 is shown in Fig.~\ref{histlc}, in terms of observed X--ray
flux in the 0.2--2~keV band. The {\it ROSAT} data are not shown for
clarity. However, as reported in Table~\ref{tab1}, the source was not
detected during the {\it ROSAT} All--Sky--Survey (RASS) and was also
undetected in pointed PSPC observations on July 1993 and June
1994. The most stringent $2\sigma$ upper limit comes from the PSPC
observation on 1994--06--29 which can be used to obtain a 0.2--2~keV
flux of $\leq 7\times 10^{-15}$~erg~s$^{-1}$~cm$^{-2}$, a factor of
$\geq$~240 lower than the observed flux during the {\it XMM--Newton}
slew $\sim 16$ years later, and a factor of $\geq$~280 lower than
during the pointed observation on 2010--12--02.  During the
$\sim$~1~yr--long monitoring which followed the {\it XMM--Newton} slew
detection, the soft X--ray flux of GSN~069 exhibits variability on all
probed timescales (down to a few hundreds of seconds, see below), but
was stable to within a factor 2 during the monitoring campaign (see
Fig.~\ref{histlc}). This allows us to exclude that the X--ray activity
is associated with a fast, transient phenomenon such as a tidal
disruption event for which a fast decline of the X--ray emission is
observed (e.g. Esquej et al.2008; Saxton et al. 2012).

\subsection{Short--timescale X--ray variability and black hole mass estimates}
\label{shortvarsec}

\begin{figure}
\begin{center}
\includegraphics[width=0.33\textwidth,height=0.45\textwidth,angle=-90]{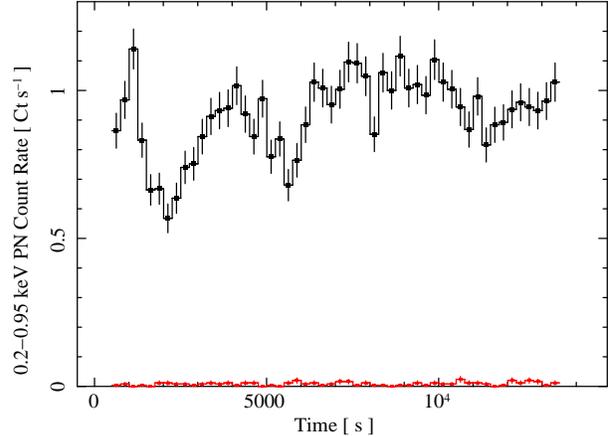}
\caption{The soft (0.2--0.95~keV) X--ray EPIC--pn light curve of GSN~069
  from the {\it XMM--Newton} pointed observation. The background light
  curve is also shown for reference after being corrected for the
  extraction area. We use a bin size of 250~s for both light curves.}
\label{lc02to1}
\end{center}
\end{figure}

The {\it XMM--Newton} pointed observation is long enough
($\sim$~15~ks) to investigate the short--timescale variability of the
source. The EPIC--pn light curve of GSN~069 in the soft
(0.2--0.95~keV) X--ray band is shown in
Fig.~\ref{lc02to1}. Significant variability is detected down to the
bin size (250~s). Since the excess variance $\sigma^2_{\rm{rms}}$ is
anti--correlated with the black hole mass (e.g. Nikolajuk, Papadakis
\& Czerny 2004), X--ray variability can be used to obtain an estimate
of the black hole mass in GSN~069. We refer to Ponti et al. (2012) for
the operative definition of $\sigma^2_{\rm{rms}}$ and its error. We
compute $\sigma^2_{\rm{rms}}$ from the 0.2--0.95~keV light curve shown
in Fig.~\ref{lc02to1} using an interval of 10~ks duration and a bin
size of 250~s. This choice corresponds to the $\sigma^2_{\rm{rms,10}}$
of Ponti et al. (2012) and we measure $\sigma^2_{\rm{rms,10}} =
3.4^{+6.6}_{-0.5}\times 10^{-2}$. As shown by Ponti et al. (2012), the
excess variances in the soft and hard X--ray bands obey a tight 1:1
correlation. Hence, we can use the derived $\sigma^2_{\rm{rms,10}}$ in
the soft band to obtain an X--ray--variability estimate of the black
hole mass in GSN~069 by using the correlations of Figure~3 in Ponti et
al. (2012), which were originally derived from the 2--10~keV light
curves. Considering the scatter in their Figure~3 rather than the
errors on their best--fitting relation, we infer a black hole mass in
the range $M_{\rm{BH,X}} \sim 0.3-7\times 10^6~M_\odot$ in GSN~069.

An independent estimate of $M_{\rm{BH}}$ can be obtained from its
relationship with the bulge K--band luminosity (Marconi \& Hunt
2003). For typical seeing conditions during 2MASS observations and
typical bulge sizes of 0.5-1~kpc, galaxy bulges are likely unresolved
beyond redshift $z\sim 0.01$. Hence, we consider the point--source
2MASS catalogue which gives $m_{\rm K} = 13.89$ ($M_{\rm K} =
-20.6$). Assuming negligible AGN contribution (see also
Section~\ref{SEDsection}), this translates into $M_{\rm{BH,K}} \leqsim
5 \times 10^6~M_\odot$, in good agreement with $M_{\rm{BH,X}}$. A
further estimate may be derived from the $M_{\rm{BH}}-\sigma_*$
relationship assuming that the Narrow--Line--Region (NLR) emitting gas
traces the stellar gravitational potential of the bulge (i.e. assuming
$\sigma_{\rm{gas}} = \sigma_*$). The relatively low resolution of our
optical spectrum only gives $\sigma_{\rm{[O~\textsc{iii}]}} \leq
120$~km~s$^{-1}$, which translates into an upper limit of
$M_{\rm{BH,\sigma_{gas}}} \leq 8.7\times 10^6~M_\odot$, also
consistent with the other estimates (we use the $M_{\rm{BH}}-\sigma_*$
relation resulting from the fits of the full sample of Xiao et
al. 2011). We conclude that all available estimates of $M_{\rm{BH}}$
consistently indicate that $M_{\rm{BH}} \simeq 10^6~M_\odot$, although
with relatively large uncertainties. As we show below, a consistent
estimate is also obtained from our UV--to--X--ray spectral modelling.

\section{The X--ray spectrum of GSN~069}
\label{secXrayspec}

The only X--ray observation with sufficient quality for detailed
spectral analysis is the pointed {\it XMM--Newton} observation
performed on 2010--12--02. The X--ray spectrum that can be obtained by
merging all {\it Swift} XRT observations contains $\sim$13 times less
X--ray counts than that from the pointed {\it XMM--Newton}
observation. We then only discuss here the spectral analysis from the
latter data set. Since fits to the individual pn and MOS spectra gave
consistent results, all instruments were fitted simultaneously,
compensating any residual normalisation difference between the EPIC
cameras with a constant. A simple power law model and Galactic
absorption (with N$_{\rm H} \equiv 2.48\times 10^{20}$~cm$^{-2}$,
Kalberla et al. 2005) cannot reproduce the data ($\chi^2 = 516$
  for 29 degrees of freedom, dof) and yields a very steep, unphysical
  photon index ($\Gamma \simeq 6.7$).

\begin{figure}
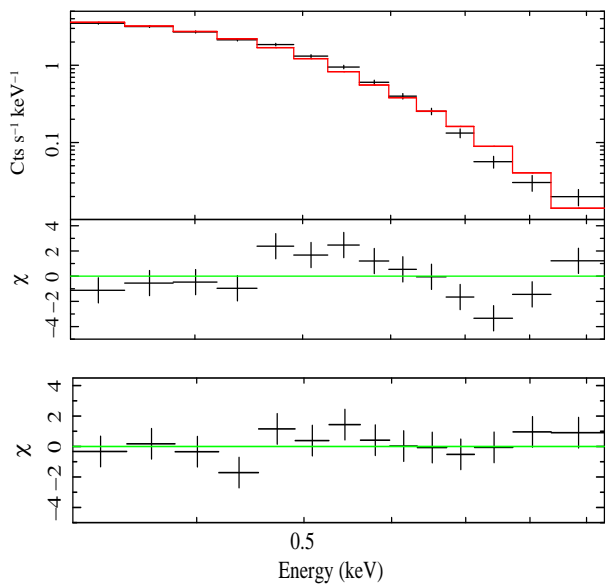

\begin{center}
\includegraphics[width=0.25\textwidth,height=0.45\textwidth,angle=-90]{spec1.ps}
{\vspace{-1.0cm}}
\includegraphics[width=0.25\textwidth,height=0.45\textwidth,angle=-90]{spec2.ps}
{\vspace{-0.0cm}}
\caption{{\bf Upper:} The EPIC pn (black) spectrum, model, and
  residuals (in terms of $\sigma$) for a simple blackbody model and
  Galactic absorption. An extra absorption feature is present around
  0.7~keV indicating the presence of ionised gas in the line of
  sight. {\bf Lower:} Best--fitting residuals obtained by adding
  a warm absorber with N$_{\rm H} \sim 1 \times
  10^{22}$~cm$^{-2}$ and $\log\xi \sim 0.40$.}
\label{specfig}
\end{center}
\end{figure}

We then replace the power law with a phenomenological (redshifted)
blackbody model in the attempt to describe the very soft spectrum. The
X--ray data are now reasonably well reproduced ($\chi^2 = 47$ for 29
dof) with a blackbody temperature of $58\pm 2$~eV. However, clear
residuals in the form of a likely absorption structure are left around
0.7~keV. In the upper panel of Fig.\ref{specfig}, we show the data,
model and residuals, showing the 0.7~keV absorption structure seen in
the data. We only show the EPIC--pn data, although the MOS ones are
also included in the fit. Adding a phenomenological edge model with
(rest--frame) energy E$_{\rm{edge}}\simeq 0.67$~keV and optical depth
$\tau\simeq 0.8$ provides a significant improvement of the fitting
statistic ($\chi^2 = 25$ for 27 dof), showing that an absorption
feature is a viable explanation of the residuals in the data, most
likely indicating the presence of partially ionised gas in the
line--of--sight. We then replace the edge with a more self--consistent
warm absorber model, namely the {\small{ZXIPCF}} model (Reeves et
al. 2008), and obtain a good description of the data ($\chi^2 = 24$
for 27 dof) for an ionised absorber with column density N$_{\rm H} =
1.0^{+0.6}_{-0.7}\times 10^{22}$~cm$^{-2}$ and ionisation parameter
$\log\xi = 0.40^{+0.09}_{-0.07}$. The blackbody temperature is now
$60\pm 2$~eV. No statistical improvement is obtained by adding a
further neutral absorption intrinsic component (i.e. at redshift
$z=0.018$) with N$_{\rm H}^{z} \leq 2\times 10^{20}$~cm~$^{-2}$. The
residuals for our final best--fitting model are shown in the lower
panel of Fig.\ref{specfig}.

 Our phenomenological blackbody model provides a fair description of
 the X--ray {\it XMM--Newton} spectrum of GSN~069, once warm
 absorption is taken into account. The inferred temperature of
   $kT\sim 60$~eV is significantly cooler than that of the typical
   soft X--ray excess in AGN which, when described in terms of thermal
   emission, is generally in the range of $100-200$~eV
   (e.g. Piconcelli et al. 2005; Crummy et al. 2006; Miniutti et
   al. 2009). The temperature derived for GSN~069 is in fact
 consistent with that expected from thermal accretion disc emission
 for a relatively small--mass black hole radiating close to its
 Eddington limit, so that the most natural interpretation for the
 spectral shape is that it represents the high--energy tail of the
 disc emission.

On the other hand, other spectral models can equally well describe the
data. In particular, the soft X--ray spectrum of GSN~069 may result
from strong Comptonization of a seed photons spectrum in a corona with
plasma temperature $kT_{\rm e}$ and optical depth $\tau_{\rm
  e}$. However, Comptonization models suffer from a well known series
of parameter degeneracies when applied to limited bandpass data. In
particular, the observed spectrum can be well reproduced by assuming a
low seed photons temperature ($\sim 10$~eV) and a corona with $kT_{\rm
  e} \sim 60$~eV and large optical depth ($\tau_{\rm e} \geqsim
10$). However, the same spectral shape can also be obtained by
increasing the seed photons temperature as well as $kT_{\rm e}$, and
by lowering $\tau_{\rm e}$ in order not to over--produce high--energy
X--rays. In summary, the properties of the putative X--ray corona are
largely unconstrained by our data. Hence, although we caution that the
spectrum may well be interpreted in terms of Comptonised emission, we
refrain from applying Comptonization models to the X--ray data here.

\subsection{A more physical description}

The best--fitting blackbody model described above
under--predicts the simultaneous UV data from the Optical Monitor (OM)
by more than two orders of magnitude\footnote{In extrapolating the
  best--fitting X--ray model, we consider UV reddening by assuming the
  standard gas--to--dust conversion $\rm{E(B-V)}=1.7\times
  10^{-22}$~N$_{\rm H}$ for the Galactic column density using the
  {\small{UVRED}} model (valid between 1000--3704~\AA): Furthermore,
  we assume that the X--ray warm absorber is dust--free, so
  does not contributes in the UV.}. Although the OM fluxes are likely
contaminated by the host galaxy emission, it seems unlikely that the
AGN does not contribute at all at least at the shortest
wavelengths. The flux in the shortest--wavelength M2 filter at $\sim
2300$~\AA\ is under--predicted by a factor of $\sim 250$ by the
best--fitting X--ray model.

We then replace the blackbody with a more physical accretion disc
model, namely the {\small{OPTXAGNF}} model (Done et al. 2012; Jin et
al. 2012) and consider simultaneous fits to the X--ray and UV
(M2--filter only) data. The main model assumptions are the following: the disc
emits a colour--corrected blackbody down to a given coronal radius
$r_{\rm c}$, and  the corona is assumed to comprise a two--phase
plasma. Within the corona, the available energy is distributed between
powering the soft X--ray excess via Comptonization in an optically
thick plasma, and the standard high--energy power law via
Comptonization in an optically thin phase.

We first assume a pure thermal accretion disc (AD) model, with no
additional Comptonization component(s). The model depends on black
hole mass, Eddington ratio, and black hole spin. We found that
statistically equivalent fits could be obtained with non-spinning and
maximally spinning black hole models.  For the case of a non--rotating
Schwarzschild black hole, the model provides a fair description of the
UV and X--ray data with a reduced $\chi^2/{\rm{dof}} = 25/28$. The
best--fitting model implies a black hole mass of $M_{\rm{BH}} =
(1.2\pm 0.1) \times 10^6~M_\odot$ and $L_{\rm{Bol}}/L_{\rm{Edd}} =
0.51\pm 0.05$. The warm absorber properties are N$_{\rm H} = (8.6\pm
1.2)\times 10^{21}$~cm$^{-2}$ and $\log\xi = 0.35 \pm 0.07$,
consistent with those derived with the simpler model discussed
above. If a maximally--rotating Kerr black hole is assumed instead,
the data are reproduced with the same statistical quality but with
larger black hole mass ($M_{\rm{BH}} \sim 8.5 \times 10^6~M_\odot$)
and lower Eddington ratio ($\sim 0.06$). We conclude that a thermal
accretion disc (AD) plus X--ray warm absorption model provides a
reasonable description of the X--ray and shortest--wavelength UV data
of GSN~069.

Next, we consider the possible presence of Comptonization in the
framework of the {\small{OPTXAGNF}} model. As no high--energy power
law is seen in the data, we assume that all the energy available to
the corona powers only the soft excess emission, which depends on the
seed photons temperature (self--consistently derived from black hole
mass and Eddington ratio) and on the coronal parameters ($kT_{\rm e}$
and $\tau_{\rm e}$) and size ($r_{\rm c}$). However, the UV data from
the M~2 filter are not sufficient to remove the degeneracies of the
Comptonization model, and we cannot constrain all the parameters
independently. We only report here results from one particular case,
chosen because such configuration is often invoked as an explanation
for the soft X--ray excess in AGN (e.g. Done et al. 2012; Jin et
al. 2012). We assume that the corona is strongly optically thick with
$\tau_{\rm e} = 13$, corresponding to the average value for the
coronal optical depth in the study of a relatively large sample of
type 1 AGN by Jin et al. (2012) who used the same spectral model
({\small{OPTXAGNF}}). The fit is statistically good
($\chi^2/{\rm{dof}} = 24/27$), and we obtain $kT_{\rm e} = 70 \pm
20$~eV for a black hole mass of $M_{\rm{BH}} = (0.9\pm 0.1) \times
10^6~M_\odot$ and $L_{\rm{Bol}}/L_{\rm{Edd}} = 0.94\pm 0.06$,
corresponding to a Bolometric luminosity of $0.9-1.3\times
10^{44}$~erg~s$^{-1}$.

As Comptonization is consistent but not required by the data, and
given that the coronal physical parameters cannot be constrained
independently, we consider that the pure AD model is the simplest
best--fitting description of the UV--to--X--ray spectrum of GSN~069,
with the caveat that we cannot exclude that the soft X--ray emission
is in fact Comptonised. The observed 0.5--2~keV flux is $\sim
1.8\times 10^{-13}$~erg~s$^{-1}$~cm$^{-2}$ yielding an unabsorbed
luminosity of $\sim 1.2\times 10^{42}$~erg~s$^{-1}$ in the same band
(we extrapolate our best--fitting up to 2~keV, and we refer to the
commonly used 0.5--2~keV band to ease the comparison with other
sources and observatories). On the other hand, the 2--10~keV
luminosity is $\leq 4.4 \times 10^{40}$~erg~s$^{-1}$. The overall
Bolometric luminosity is $\sim 8\times 10^{43}$~erg~s$^{-1}$,
corresponding to an Eddington ratio of $\sim 0.5$ and a black hole
mass of $\sim 1.2\times 10^6~M_\odot$ for a non--rotating black hole.

\section{Optical properties: GSN~069 as a true Seyfert~2 galaxy candidate}
\label{AATsection}

Two consistent 2dF and 6dF optical spectra taken in 2001 and 2003 at
the AAT show unresolved Balmer lines with no apparent broad
components. The standard diagnostics emission line ratios
[O~\textsc{iii}]$\lambda$5007~/~H$\beta \sim 9.2$ and
[N~\textsc{ii}]$\lambda$6583~/~H$\alpha \sim 1.4$ from the 2001/2003
spectra classify GSN~069 as a Seyfert galaxy, well separated from
LINERs and composite/star--forming galaxies (Baldwin, Phillips \&
Terlevich 1981; Veilleux \& Osterbrock 1987; Ho, Filippenko \& Sargent
1997). The lack of broad components indicates that GSN~069 appears to
be a rather typical Seyfert~2 galaxy in the optical at epoch
2001/2003. The dramatic X--ray brightening of the source between 1994
and 2010 prompted us to perform a new optical spectroscopic
observation with the goal of investigating any variability of the
optical spectrum and, in particular, to search for the appearance of
any broad line component. A new optical spectrum was then taken at the
AAT on 2010-10-27, $\sim$~3.5 months after the {\it XMM--Newton} slew
detection.

The spectrum has been fitted to a linear combination of spectral
templates of evolutionary star formation episodes from Bruzual \&
Charlot (2003) plus a power law to account for any AGN continuum. We
point out that replacing the arbitrary power law in the optical range
with our UV--to--X--ray best--fitting pure AD model (Section~4.1)
works very well and gives the same results. The remaining emission
lines have been fitted with Gaussian profiles in the residuals of the
stellar template fit. All can be fitted with narrow Gaussian
components, and we do not detect any broad component in any of the
emission lines down to the instrumental resolution, as was the case
also in the earlier 2001/2003 2dF and 6dF spectra. The diagnostic line
ratios are unchanged with respect to previous measurements, as
expected given the typical geometrical scale of the
Narrow--Line--Region (NLR), namely hundreds--thousands of light years.

The optical properties of GSN~069 classify it as a Seyfert
galaxy whose optical BLR is either obscured, absent, or gives rise to
undetectable broad emission lines. The lack of X--ray cold absorption
and, perhaps most importantly, the fast X--ray variability at soft
X--ray energies, rules out a standard Seyfert~2 galaxy interpretation,
so that it appears unlikely that the BLR of GSN~069 is obscured. The
remaining possibilities are that i) the BLR is present, but we are
unable to detect the corresponding broad lines in the available
optical spectra or ii) GSN~069 is a true Seyfert~2 galaxy lacking the
BLR.

\begin{figure}
\begin{center}
\includegraphics[width=0.49\textwidth,height=0.72\textwidth,angle=0]{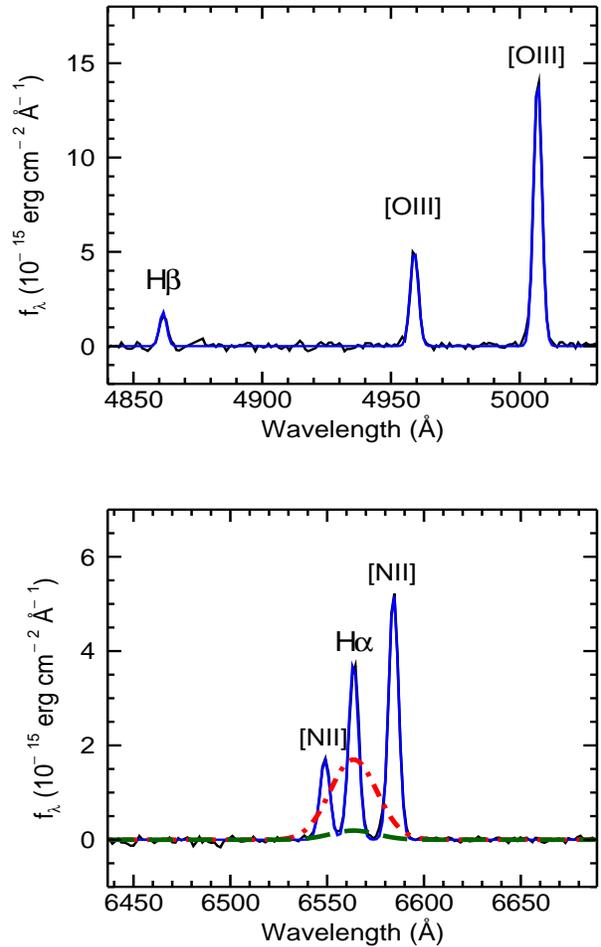}
\caption{The optical residual spectrum of GSN~069 from the
  2010--10--27 AAT observation in the H$\beta$ and [O~\textsc{iii}]
  (upper panel) and H$\alpha$ and [N~\textsc{ii}] region (lower panel)
  after subtracting the stellar contribution is shown in black. All
  emission lines have been fitted with unresolved Gaussian lines (blue
  solid line). We also show as a dotted--dashed line the expected
  broad H$\alpha$ component (see text for details). The green line is
  a broad Gaussian with the same FWHM and with intensity $\sim$~3
  times the residuals, i.e. a representation of the broad component
  upper limit. Its luminosity is $\sim$~9 times lower than that we
  would expect to be present in GSN~069.}
\label{AAT2010}
\end{center}
\end{figure}

It is instructive, at first, to estimate the FWHM and luminosity of
the broad lines that we may expect to be present in GSN~069. We focus
here on the H$\alpha$ emission line, as it is significantly stronger
than H$\beta$. From our best--fitting spectral model, we can extract
estimates of the black hole mass and intrinsic continuum monochromatic
luminosity at $5100$~\AA. By using the standard,
reverberation--mapping--calibrated relationship (e.g. Xiao et
al. 2011) between optical luminosity, line FWHM, and black hole mass,
we infer that GSN~069 should exhibit broad lines with FWHM$\sim
1300$~km~s$^{-1}$, which would classify it as a typical NLS1
galaxy. On the other hand, by using the empirical relation between
broad H$\alpha$ and continuum luminosities from Greene \& Ho (2005) as
updated in Xiao et al. (2011), we estimate
L$^{\rm{est}}_{\rm{H_\alpha}} \sim 3.7\times 10^{40}$~erg~s$^{-1}$ for
the broad component in GSN~069.

The new 2010 AAT optical spectrum is shown in Fig.~\ref{AAT2010}
around the H$\beta$--[O~\textsc{iii}] and around the
H$\alpha$--[N~\textsc{ii}] regions. We superimpose on the observed
H$\alpha$ spectrum the expected broad H$\alpha$ component, assuming
the FWHM and broad L$^{\rm{est}}_{\rm{H_\alpha}}$ estimated above. It
is quite clear that, were the broad H$\alpha$ line present, we would
have securely detected it, unless its properties (FWHM and luminosity)
were very peculiar. We must conclude that GSN~069 either lacks the BLR
or that the corresponding emission lines are unusually weak. In the
first case, GSN~069 appears to be a bona--fide true Seyfert~2 galaxy
candidate (although spectro--polarimetric observations would be needed
to exclude the presence of HBLR). 
\begin{figure*}
\begin{center}
\includegraphics[width=0.6\textwidth,height=0.85\textwidth,angle=90]{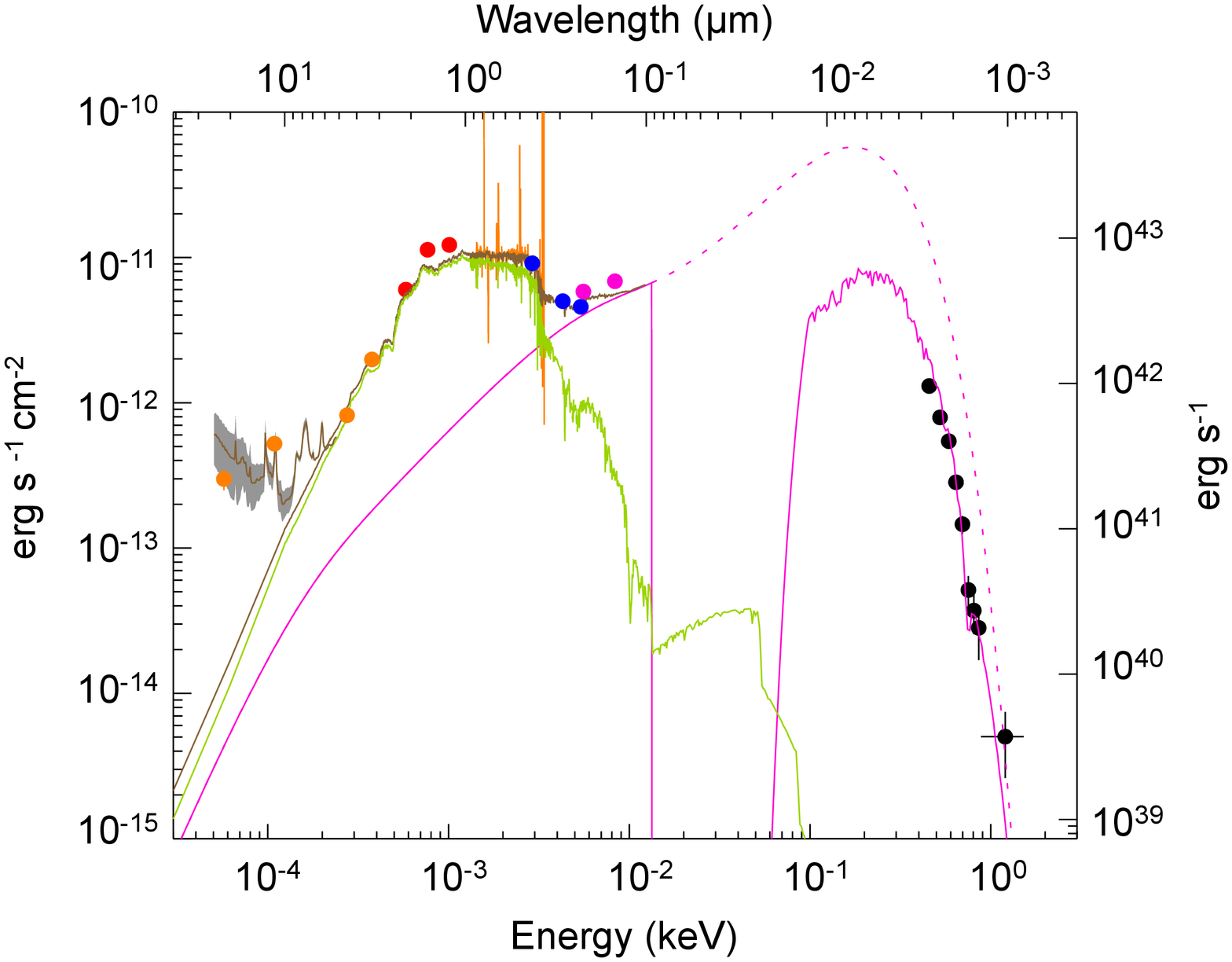}
{\vspace{-0.5cm}}
\caption{The Galactic--absorption--corrected SED of GSN~069 from
  22~$\rm{\mu m}$ to 1.5~keV, see text for details. From left to right
  we show data from {\it WISE} (orange), 2MASS (red), AAT (orange),
  {\it XMM--Newton} OM (blue), {\it GALEX} (pink), and {\it
    XMM--Newton} EPIC--pn (black). The AGN intrinsic SED is shown as a
  dotted line, while the effect of the X--ray dust--free warm absorber
  is shown as a solid line. The highest--energy X--ray data point is
  not used in the spectral analysis due to the low significance of the
  detection above 0.95~keV, but is shown here as reference. The deep
  UV/EUV trough in the warm-absorbed SED is a simple estimate of the
  absorption due to neutral hydrogen of the warm absorber in that
  spectral range. No extinction is expected below 13.6~eV under the
  reasonable assumption that the warm X--ray--absorbing gas is
  dust-free.}
\label{sed}
\end{center}
\end{figure*}

\section{Broadband SED}
\label{SEDsection}

We have collected all photometric information available for GSN~069
from various catalogues. Multi--epoch data exist in the optical B$_j$
and R bands. No significant variability is present and B$_j \simeq
16.2$ and R$\simeq 15.3$ over $\sim$~30~yr timescales\footnote{GSN~069
  is present in the USNO--B1.0 and USNO--A2.0 catalogues where
  magnitudes B$_j\sim 12-13$ and R$\sim 12$ are reported for epochs
  $\leq 1998$. However, we have analysed the corresponding digitised
  plates in the B$_j$ and R bands and we have compared GSN~069 count
  rate with those from two reference stars in all plates and in the
  {\it XMM--Newton} B--band OM observation. We do not confirm the
  magnitudes reported in those catalogues and we infer a maximum
  $60$~\% variability in the B$_j$ band between 1980 and 2010. Our
  analysis is also supported by the reported B$_j = 16.03$ obtained
  during the Durham/SAAO survey at epoch $\sim 1983.4$ by Metcalfe et
  al. (1989).}. On the other hand, some variability in the UV is
present, as the {\it GALEX} 2007 flux appears to be $\sim$~50 per cent
higher than that from the 2010 {\it XMM--Newton} OM observation.

The Galactic--absorption--corrected SED of GSN~069 is shown in
Fig.~\ref{sed} together with a plausible model. The longest wavelength
{\it WISE} data are accounted for by a Starburst template obtained
from 16 Starburst galaxies with redshift $\leq 1.3$ observed with {\it
  Spitzer}/IRS (Hern\'{a}n--Caballero \& Hatziminaoglou 2011). The
2MASS and optical data are dominated by a combination of spectral
templates of evolutionary star formation episodes with ages between
900~Myr and 5~Gyr (Bruzual \& Charlot 2003), plus a $\sim$~10 per cent
AGN contribution. The UV--to--X--ray data are well reproduced by the
pure AD SED, modified by absorption from dust--free partially ionised
gas in the X--rays (and unobservable EUV range) which provides a good
fit in the UV--to--X--ray band (Section~4.1). The discrepancy between
the UV {\it XMM--Newton} OM and {\it GALEX} data (2010 and 2007
respectively) falls into the AGN--dominated spectral region, so that
it can be naturally explained by moderate AGN variability. The gentle
rise toward shorter wavelengths of the {\it GALEX} data points
confirms the general AGN intrinsic spectral shape in that region.

\section{Discussion~I: X--ray properties and comparison with similar sources}
\label{discussionI}

 We present our discussion of the observational results in two
 steps. Here, we discuss the extreme long--term X--ray variability of
 GSN~069 and the lack of hard X--ray emission, and we compare its most
 peculiar properties with a similar, newly discovered AGN as well
 as with the class of black hole binaries. We also discuss how
 GSN~069--like sources (with more typical, larger black hole mass) may
 be missed and mis--classified in current X--ray surveys. A detailed discussion
 on the lack of BLR lines is deferred instead to
 Section~\ref{discussionII}.

The present (2010) Bolometric luminosity of GSN~069 is $\sim 8\times
10^{43}$~erg~s$^{-1}$, corresponding to an Eddington ratio $\lambda
\simeq 0.5$ for a $1.2\times10^6~M_\odot$ black hole. The luminosity
was likely similar or slightly higher in 2007, as suggested by the
{\it GALEX} fluxes (Fig.~\ref{sed}). An estimate of the past
Bolometric luminosity of GSN~069 can be obtained by applying
Bolometric corrections to the NLR emission lines. As the NLR are
likely located hundreds to thousands of light years away from the
nucleus, the NLR--based luminosity represents an estimate of the
average Bolometric luminosity that GSN~069 had in the past. Hereafter,
we refer to that estimate as the {\it historical} Bolometric
luminosity of the source. L$^{\rm{hist}}_{\rm{Bol}}$ can be estimated
from the extinction--corrected H$\alpha$ and/or [O~\textsc{iii}]
luminosity. We obtain L$^{\rm{hist}}_{\rm{Bol}} \sim 3.1\times
10^{42}$~erg~s$^{-1}$ from the H$\alpha$ luminosity with the Greene \&
Ho (2005) Bolometric correction, and a consistent
L$^{\rm{hist}}_{\rm{Bol}} \sim 2.3-5.2\times 10^{42}$~erg~s$^{-1}$ by
applying the luminosity--dependent Bolometric corrections of Lamastra
et al. (2009) or Stern \& Laor (2012) to the [O~\textsc{iii}]
luminosity. Hereafter we assume L$^{\rm{hist}}_{\rm{Bol}} \sim
3.1\times 10^{42}$~erg~s$^{-1}$, corresponding to an Eddington ratio
$\lambda^{\rm{hist}} \sim 2 \times 10^{-2}$ for a black hole mass of
M$_{\rm{BH}} \sim 1.2\times 10^6~M_\odot$. We conclude that GSN~069 is
much more luminous now than in the past with a Bolometric luminosity
$\sim$~20--30 times higher than the historical one.

\subsection{Extreme long--term X--ray variability}
\label{rosatsection}

The comparison between the {\it XMM--Newton} and {\it Swift}
observations and the {\it ROSAT} non--detection of the source implies
that GSN~069 was at least a factor of $\geq 280$ fainter in the soft
X--rays in 1994 than it is now. As our best--fitting spectral model
for the {\it XMM--Newton} pointed observation comprises significant
absorption from partially ionised gas, one possibility is that changes
in the absorber's properties are responsible for the dramatic X--ray
flux variation. For instance, the flux variability can be fully
accounted for by assuming that the absorber's column density was $\geq
8.2\times 10^{22}$~cm$^{-2}$ during the {\it ROSAT}
observation. 

 Another possibility is, however, more appealing to us. As discussed
 above L$^{\rm{hist}}_{\rm{Bol}} \sim 3.1\times 10^{42}$~erg~s$^{-1}$
 was 20--30 times lower than the present L$_{\rm{Bol}}$. If the gas
 responsible for the X--ray warm absorber has the same properties
 (density and distance) in both past low--luminosity and present
 high--luminosity states, its ionisation state would have been lower
 by a similar factor in the past, which further depresses the soft
 X--rays due to increased opacity. Indeed, if the Bolometric
 luminosity and warm absorber ionisation are both lowered by a factor
 of $\geq 15$, the expected soft X--ray flux is lower than the {\it
   ROSAT} upper limit, thus providing a clean, self--consistent
 explanation for the non--detection by {\it ROSAT}. In summary the
 extreme X--ray brightening by a factor of $\geq 280$ from 1994 to
 2010 can be explained by a much more reasonable factor of $\geq 15$
 in luminosity, consistent also with the comparison between the
 historical and present Bolometric luminosity estimates.

\subsection{On the lack of hard X--rays in GSN~069}

The intrinsic 0.5--2~keV luminosity of GSN~069 is $L_{\rm{0.5-2}}
\simeq 1.2 \times 10^{42}$~erg~s$^{-1}$, while the 2--10~keV
luminosity is $L_{\rm{2-10}} \leq 4.4\times 10^{40}$~erg~s$^{-1}$. As
shown e.g. in Miniutti et al. (2009) the 0.5--2~keV and 2--10~keV
luminosities obey a tight correlation in a sample comprising PG
quasars and AGN with small black hole mass and high Eddington ratio
(of the order of those estimated for GSN~069). According to that
correlation (see their Fig.~4), the observed upper limit on
$L_{\rm{2-10}}$ implies that the hard X--ray emission in GSN~069 is
$\geqsim 30$ times fainter than in typical type 1 AGN, even
considering AGN with similar black hole mass and Eddington ratio.

 The upper limit on the 2-10~keV luminosity translates into a 2-10~keV
 X--ray Bolometric correction $k_{2-10} \geq 1800$. The typical
 $k_{2-10}$, as obtained e.g. in Vasudevan et al. (2009) does not
 exceed $\sim 100$, i.e. it is more than one order of magnitude
 lower. GSN~069 can be classified as an (hard) X--ray weak AGN with an
 optical (2500~\AA) to X--ray (2~keV) slope $\alpha_{\rm{ox}} \leq
 -2.0$, while, based upon its intrinsic optical luminosity, we would
 expect $\alpha^{\rm{expected}}_{\rm{ox}} \sim -1.1$ (e.g. Just et
 al. 2007). In this sense, GSN~069 resembles the typical AGN in the sample of
 intermediate--mass black hole of Greene \& Ho (2004; 2007), which is
 characterised by AGN with small black hole mass and relatively high
 Eddington ratio. In a recent work, Dong, Greene \& Ho (2012) have
 shown that the $\alpha_{\rm{ox}}$ values of their sample of AGN with
 black hole masses of $10^5-10^6~M_\odot$ are systematically lower
 than the extrapolation of the well know
 $\alpha_{\rm{ox}}-l_{\rm{2500}}$ relationship derived from more
 massive systems, so that a significant fraction of AGN with low--mass
 black holes appear to be more X--ray weak than their massive
 counterparts. Some of these X--ray weak objects may be significantly
 absorbed, but absorption seems unable to explain the bulk of this
 population raising the possibility that some are intrinsically X--ray
 weak. As the Dong et al. (2012) sources are observationally biased
 towards high Eddington ratios, a connection between X--ray weakness
 and Eddington ratio seems plausible. 

In this respect GSN~069 may represent an extreme case of a
high Eddington ratio system in which the standard (optically thin)
X--ray corona is absent or unable to efficiently up-scatter the soft
disc photons. Indeed, as shown in our spectral modelling, it is quite
striking that the UV--to--X--ray data can be modelled with a pure AD
spectrum with no X--ray corona at all. Among the possible explanations
for the lack/weakness of optically thin coronal emission, we point out
the work by Proga (2005) who discusses how a failed disc--wind may
quench the coronal X--ray emission. In that situation, the relatively
high density in the failed wind means that bremsstrahlung losses
dominate over inverse Compton, thus quenching the hard X--ray
emission, while preserving the soft X--rays.

\subsection{Comparison with 2XMM~J123103.2+110648}

Terashima et al. (2012) and Ho, Kim \& Terashima (2012) report the
discovery and subsequent study of the AGN 2XMM~J123103.2+110648, which
shares many properties with GSN~069. The source is detected at soft
X--ray energies only with similar 0.5--2~keV luminosity ($\sim
2.5\times 10^{42}$~erg~s$^{-1}$) to GSN~069. The X--ray spectrum of
2XMM~J123103.2+110648 can be described by a pure AD model or by
Comptonization in an optically thick plasma, as is the case for
GSN~069\footnote{Based on X--ray flux and spectral variability
  Terashima et al. (2012) suggest that Comptonization is the most
  plausible interpretation.}. 2XMM~J123103.2+110648 is unabsorbed in
the X--rays and does not show any broad Balmer lines in the optical,
strongly suggesting that this AGN lacks the BLR too (or that the
corresponding emission lines are much weaker than in typical AGN).

In summary, GSN~069 and 2XMM~J123103.2+110648 are bona--fide true
Seyfert~2 galaxies with relatively small black hole mass
($10^5-10^6~M_\odot$) and high Eddington ratio ($\sim 0.5$). Their
soft X--ray emission is variable on both short and long timescales,
while no X--ray photons are securely detected above 1--2~keV. Besides the
lower temperature of the soft excess in GSN~069 (which may reflect a
slightly higher black hole mass), the only apparent difference between
the two sources is that GSN~069 appears to be more luminous now than
in the past with ${\rm{ L^{present}_{Bol} / L^{hist}_{Bol} }} \sim
20-30$, while 2XMM~J123103.2+110648 does not appear to have
dramatically changed its radiative output, i.e.  ${\rm{
    L^{present}_{Bol} / L^{hist}_{Bol} }} \sim 1-2$ (Terashima et
al. 2012; Ho et al. 2012).

\subsection{GSN~069 in the wider context: X--ray binaries/AGN unification and  Compton--thick AGN population}

It is interesting to compare the peculiar SED of GSN~069 with that of
black hole X--ray binaries. During their outbursts, these systems
reach a so--called soft or thermal state that is dominated by thermal
AD emission in the X--ray band. During the thermal state, a power law
component is generally seen, with a fractional contribution of
$\leqsim 20$ per cent (Remillard \& McClintock 2006; Dunn et
al. 2010). Some X--ray transients, however, do reach super--soft
states in which the hard X--ray contribution appears to be
negligible. Here we only mention the case of XTE~J1650--500 which,
during its 2001/2002 outburst, reached a super--soft state with a
$\leqsim 0.5$ per cent power law contribution to the Bolometric
luminosity (Motta et al. in preparation; private communication), as it
is the case in GSN~069. The super--soft state in XTE~J1650--500 lasted
about one month which, accounting for a linear scaling between
timescales and black hole mass, would translate into the possibility
that the super--soft state in GSN~069 may last as long as a few
thousands of years. Considering that the overall outburst of
XTE~J1650--500 was about 6 months long, the duty cycle for such
super--soft state may typically be of the order of 10--20 per cent in
accreting black holes with Eddington ratios above ${\rm{few}}\times
10^{-3}$. A study of the occurrence of these states in the known X--ray
binary population will enable us to refine the above estimates.

On the other hand, variability properties define a striking difference
between GSN~069 and black hole X--ray binaries in soft, thermal
dominated states. GSN~069 exhibits short--timescale variability on
timescales as short as a few hundreds of seconds (Fig.~\ref{lc02to1}),
while the AD emission in black hole binaries is stable on timescales
$\leqsim 1$~s (corresponding to $\leqsim 10^5$~s for a $10^6~M_\odot$
black hole as in GSN~069). This may suggest that the soft X--ray in
GSN~069 are Comptonised rather than purely thermal (as suggested by
Terashima et al. 2012 for 2XMM~J123103.2+110648) or that
the AD structure in AGN is slightly different from that in X--ray
binaries (maybe inhomogeneous and intrinsically more variable).

Finally, it is interesting to consider the observational analogy
between GSN~069 and 2XMM~J123103.2+110648 and heavily absorbed,
Compton--thick type~2 AGN candidates. Both classes of sources are
characterised by relatively strong narrow optical lines such as
O~\textsc{iii}, the absence of broad optical lines, and the lack of
detection in the hard X--rays (or, in any case, a high
$L_{\rm{O~\textsc{iii}}}/L_{\rm{X}}$ ratio). The only difference
between the two classes is that GSN~069 and 2XMM~J123103.2+110648 are
detected and variable (on short timescales) in the soft X--rays, while
the soft X--ray emission in Compton--thick AGN is related to
reprocessing in large--scale photo--ionised plasma which is, by
definition, weak and constant. If our interpretation of the soft
X--ray emission in terms of AD thermal emission is valid, the
detection of any soft X--ray emission in GSN~069 and
2XMM~J123103.2+110648 is only possible thanks to their low black hole
mass and relatively high accretion rate. More massive AGN with similar
or smaller mass accretion rates have AD emission whose high--energy
tail does not enter at all the soft X--ray band, so that large black
hole mass analogs of GSN~069 will be missed in the X--rays. In fact,
assuming a typical Eddington ratio $\lambda=0.1$ and non--rotating
black holes, the AD emission does not reach the soft X--ray band for
${\rm{M_{BH}}}\geq 10^7~M_\odot$ at low redshift, so that almost all
GSN~069--like AGN would be mis--classified as Compton--thick
candidates in even moderately high--redshift surveys because of their
relatively strong O~\textsc{iii} emission and lack of detection at
X--ray energies.

Hence, there could be an entire population of AGN in super--soft
states (in analogy with X--ray binaries) that are completely missed in
current X--ray surveys. The time that AGN spend in super--soft states
may be as long as $10^{4}-10^{5}$ years for typical black hole masses
of $10^7-10^8~M_\odot$ with a typical duty cycle of 10--20 per cent
for Eddington ratios above ${\rm{few}}\times 10^{-3}$, suggesting that
the population of super--soft state AGN may be non--negligible. If so,
GSN~069--like objects may contaminate the derived fraction of highly
obscured AGN in the Universe because, as discussed above, they are
likely counted as Compton--thick AGN in current X--ray surveys.

\section{Discussion~II: On the lack of broad emission lines in GSN~069}
\label{discussionII}

Although spectro--polarimetric observations would be needed to
conclude that GSN~069 and 2XMM~J123103.2+110648 do not have HBLR, the
lack of broad optical lines and of X--ray cold absorption, together
with the observed short timescale X--ray variability in the soft band,
suggests that both AGN may have intrinsically very weak (or absent)
BLR optical lines. Here we critically consider possible explanations
for the lack of BLR in these objects in the framework of two popular
BLR models.

\subsection{Linking the lack of hard X--rays with the absence of the BLR}

Both GSN~069 and 2XMM~J123103.2+110648 are outstanding objects not
only based on their optical and soft X--ray properties, but also
because they lack hard X--ray emission, which has up to now been
considered a hallmark property of accreting supermassive black
holes. The lack of hard X--ray emission may be invoked to account
naturally for the lack of broad optical emission lines if the BLR
clouds need to be confined and in pressure equilibrium with a hot
medium. Such two--phase BLR model was first proposed by Mathews (1974)
and then refined by e.g. Wolfe (1974), McKee \& Tarter (1975) and, in
much more detail, by Krolik, McKee \& Tarter (1981).

In particular, Krolik et al. (1981) show that, depending on the
details of the radiation field, cold, dense gas (i.e. the BLR clouds)
can coexist in pressure equilibrium with - and be confined by -
hotter, less dense gas (typically with temperatures of $\sim 10^8$~K)
at ionization stages consistent with that of the BLR. The state of the
gas is determined by its thermal (and ionisation) equilibrium with the
irradiating radiation field. At the typical distances of the BLR, the
temperature is roughly governed by the balance between Compton heating
(by X--rays) and inverse--Compton cooling (see however Krolik et
al. 1981 for a comprehensive study of other possible heating and
cooling mechanisms). As clear, the predicted temperature sensitively
depend on the hard X--ray flux which determines the amount of
Compton--heating, without affecting so strongly the cooling which
depends more on the overall spectrum. Krolik et al. (1981) show indeed
that the extent of the multi--phase region where cold BLR clouds are
in pressure equilibrium with the hotter inter--cloud medium at the
right ionisation state is a monotonic function of the fraction of high
energy photons in the radiation field. When hard X--rays are weak or
even absent, the multi--phase character of the equilibrium solutions
is lost, and cold, dense clouds cannot be confined by the hotter
medium (see also Guilbert, McCray \& Fabian 1983).

The two--phase model for the BLR thus predicts the disappearance of
the BLR emission lines whenever the inter--cloud medium is not hot
enough to support the colder/denser BLR clouds. As the Compton temperature
tends towards the SED peak in $\nu L_\nu$, one can visually estimate
that it is roughly $\sim 10^6$~K in GSN~069 (see Fig.~\ref{sed}). Such
temperature is nearly two orders of magnitude lower than that
required to confine the cold line--emitting clouds in the BLR. It is
then quite clear that the lack of hard X--rays in GSN~069 (and in
2XMM~J123103.2+110648) corresponds to a case in which the Compton
temperature is too low to sustain a multi--phase medium, thus
preventing thermodynamically the formation of a broad--line emitting
region. 

More examples of GSN~069--like objects with variable levels of hard
X--ray emission would be crucial to confirm or discard this idea in
the future (possible examples of hard--X--ray emitters lacking the BLR
may be Mrk~273x and 1ES~1927+654, as discussed by Bianchi et
al. 2012). We will continue to monitor GSN~069 in the X--rays in the
future. The detection of hard X--rays sometime in the future will
prompt us to perform another optical/UV spectroscopic campaign on this
peculiar object, which may be used to confirm/dismiss the above
interpretation, and/or to assess the relevant heating/cooling
timescales associated with the two--phase BLR model observationally.

On the other hand, problems with the two--phase model discussed above
have been subsequently recognised (see e.g. Mathews \& Ferland
1987). In particular, the Compton temperature as derived from current
detailed AGN SED is typically $10^7$~K, too low from the BLR clouds to
be stable against drag forces, or for the hot phase to be optically
thin to X--ray radiation.  Mathews \& Ferland also point out some
problems with the dynamics of the BLR clouds in the hot inter--cloud
medium and conclude that the two--phase picture for the BLR has to be
modified to be consistent with the observed emission line and
radiation field properties of AGN. Some of the problems of the
two--phase model may be solved by considering that BLR clouds are part
of an outflow (see e.g. Elvis 2000). An outflowing warm wind (with
typical temperature of $10^6$~K) with embedded cooler clouds does not
suffer from the shear stress problems of fast--moving clouds in a
stationary hot atmosphere (typical of the two--phase model), and a
geometrically thin wind also largely avoids the Compton depth problem pointed
out by Mathews \& Ferland (1987). The structure of the Krolik et
al. (1981) multi--phase medium is then retained rather than abandoned,
but many of the problems arising from the presence of an extended,
static, hot confining medium are solved quite naturally mostly thanks
to the dynamical nature of the outflowing solution. Hence, outflow
models for the BLR have received considerable attention in the last
decade. Below, we discuss the implications of our observational
results on GSN~069 for outflow--based models of the BLR in AGN.

\subsection{The BLR as part of a disc--wind}

The inability of some AGN to form the BLR is theoretically predicted
by several models, mostly based on the idea that the BLR is part of an
outflow, or disc--wind (Emmerging et al. 1992, Murray et al. 1995;
Elvis 2000; Nicastro 2000; Elitzur \& Ho 2009 and references
therein). We discuss here the implications of our observational data
for models associating the BLR with AGN outflows. In the discussion
below, we assume the best--fitting parameters of our pure AD model,
i.e. M$_{\rm{BH}} = 1.2\times 10^6~M_\odot$ and $\lambda =
0.5$. Moreover we assume the standard accretion efficiency $\eta =
0.06$ for a non--rotating black hole, and a viscosity parameter
$\alpha=0.1$.

For the above parameters, all available models predict that GSN~069
should have developed a disc--outflow already. If the BLR are
associated to disc--winds, broad emission lines should then be
present, contrary to our optical spectroscopic observations. On the
other hand, $\lambda^{\rm{hist}} \sim 2 \times 10^{-2}$, as inferred
from the NLR line luminosity. Such lower Eddington ratio may be
insufficient to give rise to a disc--wind and, moreover, it may be
associated with a different accretion flow geometry comprising, for
instance, a radiatively inefficient inner part (hereafter called RIAF
for simplicity, e.g. Begelman et al. 1984; Narayan et al. 1995; Yuan
2007), and a standard outer disc. In fact, most models predict that
the transition between a disc--RIAF solution and a disc--only one
occurs around $\lambda\simeq 10^{-2}$ (or even above, see e.g. Rozanska
\& Czerny 2000, where the transition to an inner RIAF solution is
predicted to occur at $\sim 7\times 10^{-2}$ in AGN for $\alpha\simeq
0.1$). If GSN~069 was in the past characterised by a RIAF--disc flow,
a standard disc--wind may have developed only recently, following a
relatively recent re--activation of GSN~069, i.e. a transition
from a radiatively inefficient RIAF--disc to a radiatively efficient
disc--only solution, only the latter being associated with a disc
wind.

\subsubsection{BLR formation timescale in disc wind models}

The state transition from an RIAF--disc to a disc--only
accretion mode must have occurred some time before 2007, as the {\it
  GALEX} data indicate a high radiative efficiency at that epoch (see
Fig.~\ref{sed}). This would imply that, if the BLR forms via a disc
wind, its formation timescale is longer than $\sim 3$~yr, i.e. the
time--span between the {\it GALEX} observation in 2007, when
GSN~069 already had a high $\lambda \geqsim 0.5$ (see Fig.~\ref{sed}),
and the AAT optical spectrum in 2010, where no evidence for a
BLR is found. Is a $\sim 3$~yr minimum BLR formation timescale
consistent with a disc wind origin of the BLR? Once the
launching/accelerating conditions are met, an outflow should be
launched instantly (light--travel time affects are totally negligible
here), so the answer appears to be negative. 

Let us, however, consider a general geometry of radiatively--driven
disc--wind models in some more detail. The wind is expected to
initially rise vertically at $R_{\rm{wind}}$ with velocity $v_0$. As
shown by Risaliti \& Elvis (2010) and Nomura et al. (2013) based on
non--hydrodynamical models, the wind likely comprises three main
zones: an inner failed wind which is too ionized (or too dense in the
Nomura et al 2013 study) to be efficiently radiatively accelerated; a
middle zone (typically with launching radii of the order of hundreds of $r_g$)
where ionisation is relatively low so that the gas is efficiently
radiatively accelerated up to escape velocity; an outer zone where the
local disc luminosity and the UV irradiation from the innermost
regions are weak, so that the wind never reaches escape velocity and
fails (or it is not even launched vertically). 

According to this simple geometry, the inner zone represents the
so--called shielding gas (e.g. Murray et al. 1995) which reduces the
ionisation of the outer zones; the middle zone gives rise to the
actual disc wind, which is observationally identified with the broad,
blueshifted high--ionisation--emission--lines (HIEL) such as
e.g. C~\textsc{iv}, and with the corresponding absorption--lines
whenever our line--of--sight (LOS) crosses the wind itself as in
narrow/broad--absorption--lines AGN. On the other hand, the outermost
failed wind may be responsible for some of the lower--ionisation broad
emission lines (e.g. the Blamer lines as well as Mg~\textsc{ii}) which
are generally symmetric and centred at rest--frame wavelength. The
observational evidence that the low--ionisation--emission--lines
(LIEL) see a continuum that has been filtered through the wind
(e.g. Leighly 2004; Leighly \& Moore 2004) provides further support to
this geometry for the BLR, which is often referred to as {\it{the
    wind--disc geometry of the BLR}}, to differentiate the two main
BLR regions, of which one is dominated by outflows, the other by
gravity (e.g. Richards et al. 2011 and references therein).

As our AAT observation only could target the LIEL, we are here
concerned with the so--called disc--component of the BLR only. In
order for the gas to produce the LIEL, substantial EUV and soft
X--rays irradiation is required. However, all disc wind models
(e.g. Risaliti \& Elvis 2010, Nomura et al. 2013) and simulations
(e.g. Proga \& Kallmann 2004; see also subsequent work by Schurch \&
Done 2009 and by Sim et al. 2010) demonstrate that soft X-rays and EUV
irradiation is weak close to the disc plane at relatively large radii
because of absorption and Compton scattering in the inner wind/shield,
and because of the angular dependence of the disc emission itself. In
fact, one can qualitatively infer that zones within $i_{\rm{crit}}\sim
25^\circ$ from the disc plane are unlikely to be efficiently
irradiated. Hence, any wind (either accelerated or failed) has to
initially rise vertically to a height $z\sim \tan
(i_{\rm{crit}})~R_{\rm{wind}}\sim 0.5~R_{\rm{wind}}$ in order to be
exposed to sufficient irradiation to produce the LIEL. The rising time
is obviously $\Delta t \sim 0.5~R_{\rm{wind}}~v_0^{-1}$, where $v_0$
is the vertical rising velocity.

We can then estimate the minimum launching radius
$R_{\rm{wind}}^{\rm{(min)}}$ that is consistent with the
observationally--derived minimum LIEL--BLR formation timescale of
$\sim 3$~yr for GSN~069. Assuming, for simplicity, that $v_0 =
10^2$~km~s$^{-1}$ at all radii (Risaliti \& Elvis 2010),
$R_{\rm{wind}}^{\rm{(min)}}\sim 1.9\times 10^{15}$~cm$\sim
10^4~r_g$. If the wind is launched from $R_{\rm{wind}} \geqsim
R_{\rm{wind}}^{\rm{(min)}}$ the LIEL--BLR formation timescale is
longer than $\sim 3$~yr and thus consistent with the GSN~069 data.  It
is interesting to compare this value with the BLR radius, as derived
from the BLR size--luminosity relation (e.g. Kaspi et al. 2005; Bentz
et al. 2006) which suggests $R_{\rm{BLR}} \sim 1.2\times 10^{16}$~cm
in GSN~069. Reassuringly, $R_{\rm{BLR}}$ is significantly larger than
$R_{\rm{wind}}^{\rm{(min)}}$.

We point out here that $R_{\rm{wind}}^{\rm{(min)}}\sim 10^4~r_g$ is
much larger than the typical radius where disc winds can be launched
vertically due to the local radiation pressure (see e.g. Nomura et
al. 2013). Moreover, any disc wind launched and accelerated within
$\sim 500~r_g$ to escape velocity, would reach $R_{\rm{BLR}}$ in less
than a year, thus predicting much shorter LIEL--BLR formation
timescales than observed ($\geqsim 3$~yr). We can conclude that the
LIEL are not formed directly in the standard disc wind launched at
hundreds of $r_g$ from the central black hole. 

Our results are instead consistent with the idea proposed by Czerny \&
Hryniewicz (2011) in which the LIEL--BLR originate in a
dust--driven--outflow launched off the disc between the radius
$R_{\rm{1000~K}}$ where the local temperature is $\sim 1000$~K (i.e.
dust can survive within the disc), and the dust sublimation radius
$R_{\rm{dust }}$. Czerny \& Hryniewicz (2011) show that
$R_{\rm{1000~K}} \simeq R_{\rm{BLR}}$ in objects where $R_{\rm{BLR}}$
has been measured via reverberation--mapping, so that their idea has a
sound observational basis. The presence of dust in this range of radii
boosts the effect of radiation pressure with respect to the case of a
dust--free gas, and contributes to drive a massive vertical outflow
off the disc. When the gas reaches sufficiently high elevation to be
irradiated, it looses any dust content and the outflow fails (as at
these large radii irradiation is not strong enough to drive a radial
wind, see e.g. Risaliti \& Elvis 2010). Such dynamics, comprising
rising and falling clumps, provides a turbulent medium dominated by
Keplerian motion, an elegant description of what the LIEL--BLR may
indeed look like. Moreover, the outer parts of the dust--driven--wind
beyond the dust sublimation radius may give rise to the so--called
obscuring torus which is likely outflowing due to the radiation
pressure on the dust grains, linking the BLR and the torus structures
via a unique dust--driven--outflow.

Within this model, a dust--driven--wind rising vertically from
$R_{\rm{BLR}}\simeq R_{\rm{1000~K}}$ with velocity $v_0\sim
10^2$~km~s$^{-1}$ would be irradiated after $\sim 20$~yr in GSN~069,
hence the LIEL will start to form within about 20 years of the
re--activation of GSN~069. However, for the LIEL lines to be detected,
the irradiated gas needs to reach a substantial filling
factor. Assuming that a standard BLR comprises the whole region
between $R_{\rm{1000~K}}$ and $R_{\rm{dust}}$, a fully mature
LIEL--BLR in GSN~069 may take a few hundreds of years to develop,
although weak LIEL may start to be visible a few tens of years
after the re--activation.

\subsubsection{On the wind component of the BLR}

\begin{figure*}
\begin{center}
\includegraphics[width=0.67\textwidth,height=1.0\textwidth,angle=-90]{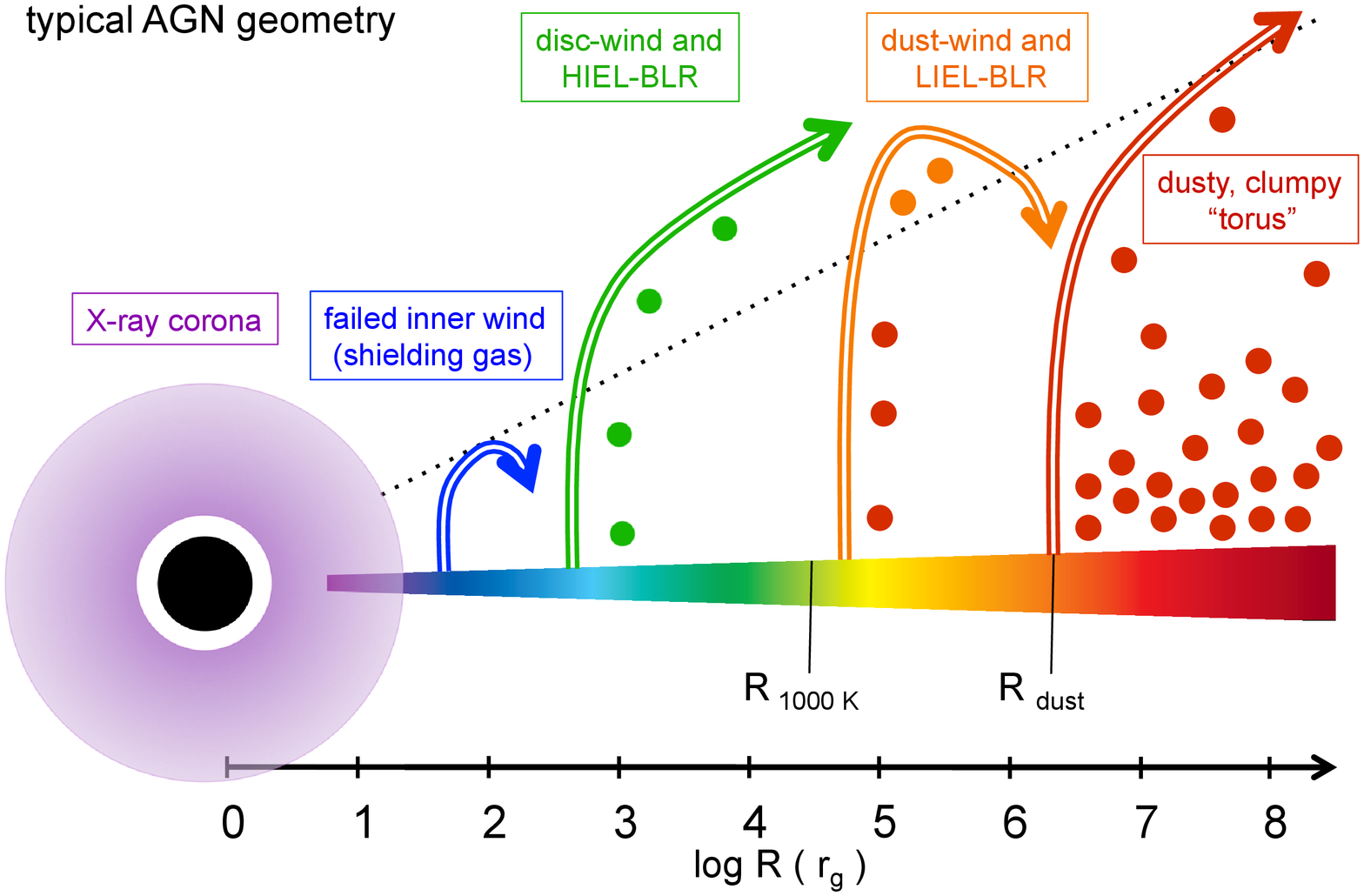}
{\vspace{-1.75cm}}
\includegraphics[width=0.67\textwidth,height=1.0\textwidth,angle=-90]{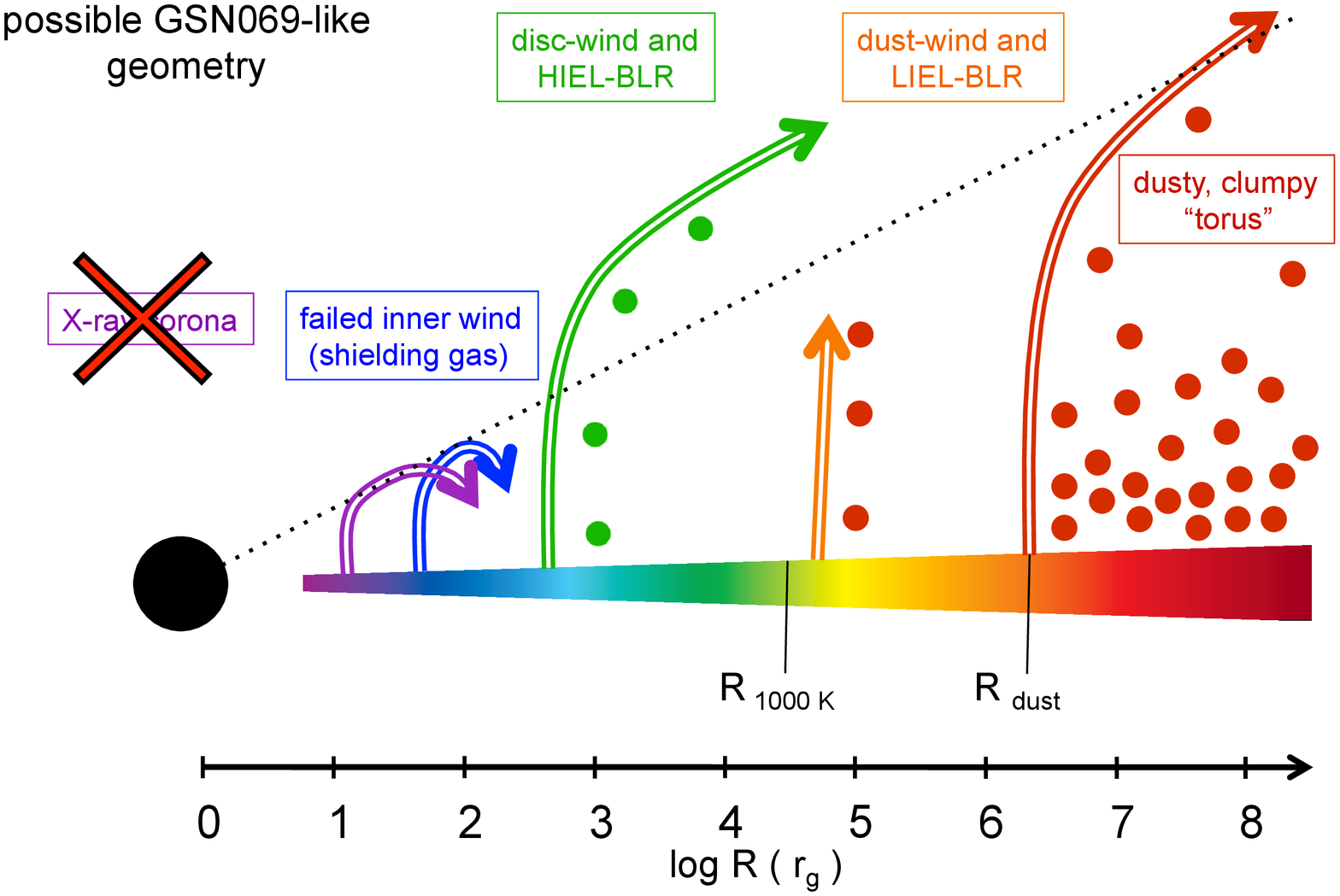}
{\vspace{-0.7cm}}
\caption{{\bf In the upper panel}, we present a schematic view of the
  possible disc--wind/BLR/torus typical geometry. The dotted line
  shows the radial direction with $i_{\rm{crit}} = 25^\circ$ from the
  disc plane. We assume that shielding is provided by an inner
  failed--wind. A standard disc--wind launched at hundreds of $r_g$
  produces the HIEL, while a failed dust--driven--wind is responsible for the
  LIEL at larger radii. {\bf In the lower panel}, we show one possible
  geometry for GSN~069. The inner failed--wind (i.e. the shielding
  gas) may extend to inner radii disrupting the corona. The
  dust--driven--wind is not yet irradiated, its motion is purely
  vertical, and no LIEL--BLR is formed yet. Whether a disc--wind
  producing the HIEL is already present could be clarified with future
  UV spectroscopic observations.}
\label{BLRgeometry}
\end{center}
\end{figure*}

As discussed above, the lack of broad LIEL in GSN~069 is consistent
with an outflow origin, provided that the wind is launched further
away than $R_{\rm{wind}}^{\rm{(min)}}\sim 1.9\times 10^{15}$, possibly
indicating a dust--driven--wind origin for the disc component of the
BLR, as suggested by Czerny \& Hryniewicz (2011). However, an inner
disc--wind should have been launched in GSN~069 already, giving rise to
potentially observable broad and blueshifted HIEL such as
e.g. C~\textsc{iv}. Future spectroscopic UV observations of GSN~069
may then reveal the presence of such a wind component which, being
launched from smaller radii, is already exposed to UV irradiation and
should reveal itself as a HIEL--emitter. Observations targeting
simultaneously both the wind (C~\textsc{iv}) and the disc
(Mg~\textsc{ii}) component of the BLR may therefore be crucial for our
understanding of the overall (wind and disc) BLR formation mechanisms
and timescales. A simple sketch of the envisaged geometry of the
system within the disc--wind (and dust--driven wind) scenario is
presented in Fig.~\ref{BLRgeometry}.

\section{Summary and conclusions}

GSN~069 can be classified as a Seyfert~2 galaxy in the optical. The
lack of cold X--ray absorption as well as the soft X--rays short
timescale variability rule out a standard Seyfert~2 galaxy
interpretation of the X--ray data, suggesting that GSN~069 is a true
Seyfert~2 galaxy candidate which lacks the BLR (or has very weak
lines). The source is undetected above 1~keV, suggesting a
particularly weak (or absent) optically thin standard X--ray corona
resulting in a very large 2--10~keV Bolometric correction $\geq
1800$. The X--ray spectrum is affected by warm absorption and it is
significantly softer than the typical AGN soft excess. It can be
described by pure AD thermal emission suggesting a black hole mass of
$\sim 1.2\times 10^6~M_\odot$ and Eddington ratio $\lambda \sim 0.5$
($L_{\rm{Bol}}\sim 8\times 10^{43}$~erg~s$^{-1}$ ). The historical
Bolometric luminosity, as traced by the NLR emitting gas, is instead
$\sim 3\times 10^{42}$~erg~s$^{-1}$, thus corresponding to a
historical Eddington ratio of $\lambda^{\rm{hist}}\sim 2 \times
10^{-2}$. Although spectro--polarimetric observations would be needed
to exclude the presence of a HBLR, GSN~069 appears to be a puzzling
true Seyfert~2 galaxy candidate with higher--than--critical Eddington
ratio for the BLR disappearance (i.e. for the disc wind disappearance
according to Nicastro et al. 2000; Trump et al. 2011). We propose two
possible scenarios that can explain our observations.

In the first scenario, the lack of optical broad lines in GSN~069 is
directly linked with the lack of hard X--rays. This is because, if the
BLR consists of relatively cold clouds in pressure equilibrium with a
much hotter inter--cloud medium, the lack of hard X--ray emission
implies a dramatic drop of the Compton temperature of the gas,
destroying the two--phase stability regime of the cloud/inter--cloud
medium, Hence, although the EUV and soft X--rays luminosity is high
enough to produce the optical emission lines, the extreme hard--X--ray
weakness of GSN~069 prevents the BLR from reaching the necessary
two--phase equilibrium, leading to the suppression of the
line--emitting region thermodynamically. Such idea may be tested by
following the evolution (if any) of the hard X--ray emission in
GSN~069 and similar objects (e.g.2XMM~J123103.2+110648 which has
similar properties and also lacks hard X--ray emission). The detection
of sources with a standard level of hard X--ray emission but with no
BLR would challenge the idea of a two--phase component for the BLR,
and would imply that this interpretation is unlikely to be viable in
GSN~069. Bianchi et al. (2012) point out the existence of two such
candidates (Mrk~273x and 1ES~1927+654) which are surely worth further
study.

In the second scenario, GSN~069 has recently experienced a transition
from a relatively radiatively inefficient state
($\lambda^{\rm{hist}}\sim 2 \times 10^{-2}$) to a highly efficient one
($\lambda\sim 0.5$, likely already in place during the 2007
{\it{GALEX}} observation). A disc--wind has then formed recently
following this re--activation, and its signature may be revealed by
looking for broad and blueshifted HIEL (e.g. C~\textsc{iv}) in future
UV spectroscopic observations. As for the LIEL, our data can be used
to infer a minimum LIEL--BLR formation timescale of $\geq$~3~yr which
is consistent with the idea of a dust--driven--wind launched at
$R_{\rm{wind}}\geqsim 1.9\times 10^{15}$~cm (consistent with the
estimated $R_{\rm{BLR}}\sim 1.2\times 10^{16}$~cm). The dust--driven
outflow is then still in its rising phase, so that no LIEL are formed
yet. The lines form when the dusty--wind begins to be irradiated (a
few tens of years after it has been launched), although it will take
about one order of magnitude more time to form a fully mature
LIEL--BLR filling the entire BLR region out to the dust sublimation
radius. On the other hand, our results seem to rule out that the LIEL
are part of a standard disc--wind launched at few hundreds of $r_g$,
as the wind would already have reached the BLR typical location
producing broad optical lines, in contrast with the minimum $\sim
3$~yr timescale we derive observationally. Within the latter,
outflow--based scenario, the lack of hard X--ray emission in GSN~069
(and 2XMM~J123103.2+110648) could be associated with a very early
phase in the transition from a radiatively inefficient to an efficient
flow. It is possible that an inner failed wind is present at this
evolutionary stage, quenching the formation of a standard X--ray
corona which may settle down into its optically--thin configuration
later on (e.g. Proga 2005).

It is also worth mentioning that the model of episodic BLR formation
proposed by Wang et al. (2012) in the context of star--forming,
self--gravitating discs in AGN predicts the existence of a (rare)
population of so--called Panda AGN that should exhibit relatively high
Eddington ratio with no BLR signatures during the early stages of the
BLR formation. Although the lifetime of such a phase may be very short
in GSN~069 (primarily due to the low black hole mass), it is
interesting to note that GSN~069 may represent one Panda AGN
candidate, to be confirmed with future spectro--polarimetric,
as well as UV spectroscopic observations.

Finally, it is interesting to note that GSN~069 (and
2XMM~J123103.2+110648) may represent (one of) the long--sought missing
links between black hole X--ray binaries and supermassive accreting
black holes. Indeed, the properties of GSN~069 suggest that this AGN
is currently in a super--soft state, similar to that experienced by
some black hole X--ray transients during their outburst evolution,
when the emission is completely dominated by the AD thermal emission,
and the hard X--ray contribution to the Bolometric luminosity is
negligible. GSN~069 was here detected and classified properly only
thanks to its small black hole mass and high Eddington ratio, which
shifts the AD thermal emission up to the soft X--rays. AGN powered by
larger black holes and/or accreting at lower rates would not have been
detected in the soft X--rays at all. It is then possible that there is
an entire population of AGN with more typical (larger) black hole mass
in super--soft states that are completely missed in current X--ray
surveys. A GSN~069--like object with larger black hole mass would in
fact be likely classified as a Compton--thick candidate. Assuming
typical Eddington ratio $\lambda=0.1$ (and non--rotating black holes),
no soft X--rays are expected from the AD thermal emission for
${\rm{M_{BH}}}\geq {\rm{few}}\times10^7~M_\odot$ at redshift zero, so
that almost all GSN~069--like AGN would be mis--classified as
Compton--thick type~2 candidates in even moderately high redshift
surveys. If the population of GSN~069--like objects with typical
$10^7-10^8~M_\odot$ black hole mass is not negligible, these sources
may then contaminate the fraction of heavily obscured AGN in the
Universe as derived from current X--ray surveys. Moreover, if the
analogy with black hole binaries holds, the life--time of super--soft
states in AGN could be as long as $10^4-10^5$~yr, i.e. orders of
magnitude longer than the time X--ray observatories have existed. As
such, we cannot exclude that the fraction of GSN~069--like,
super--soft AGN is not negligible in the Universe. Future studies will
be devoted to assess the relevance of GSN~069--like objects for the
overall AGN population.

\section*{Acknowledgements}

Based on observations obtained with XMM-Newton, an ESA science mission
with instruments and contributions directly funded by ESA Member
States and NASA. This work made use of data supplied by the UK Swift
Science Data Centre at the University of Leicester. This publication
makes use of data products from the Wide-field Infrared Survey
Explorer (WISE), which is a joint project of the University of
California, Los Angeles, and the Jet Propulsion Laboratory/California
Institute of Technology, funded by the National Aeronautics and Space
Administration. We also made use of data from the NASA Galaxy
Evolution Explorer (GALEX), operated for NASA by the California
Institute of Technology under NASA contract NAS5-98034, as well as of
data from the Two Micron All Sky Survey (2MASS), a joint project of
the University of Massachusetts and the Infrared Processing and
Analysis Center/California Institute of Technology, funded by the
National Aeronautics and Space Administration and the National Science
Foundation. We thank the anonymous referee for constructive criticism
and suggestions that significantly helped us to improve our paper. GM
also thanks Margherita Giustini and Sara Motta for valuable
discussions. GM and PE acknowledge support from the Spanish Plan
Nacional de Astronom\'{\i}a y Astrof\'{\i}sica under grants
AYA2010-21490-C02-02 and AYA2009-05705-E respectively. AMR
acknowledges UKSA funding support.

\end{document}